\documentclass[twoside]{article}
\usepackage{fleqn,espcrc2}
\usepackage{graphicx}

\newcommand{\AmS}{{\protect\the\textfont2
    A\kern-.1667em\lower.5ex\hbox{M}\kern-.125emS}}										
\def\beq{\begin{equation}}
\def\eeq{\end{equation}}
\def\bea{\begin{eqnarray}}
\def\eea{\end{eqnarray}}
\def\bq{\begin{quote}}
\def\eq{\end{quote}}

\def\ga{\left(}
\def\dr{\right)}

\def\la{\langle}
\def\ra{\rangle}
\def\nin{\noindent}
\def\ba{\begin{array}}
\def\ea{\end{array}}

\def\b{$\bullet~$}
\def\als{\alpha_s}

\def\gg2{ \la\alpha_s G^2 \ra}
\def\gg3{g^3f_{abc}\la G^aG^bG^c \ra}
\def\ggg4{\la\als^2G^4\ra}

\title
{\bf{\boldmath
{\Large  
Can one measure $C$-odd asymmetry
in  $e^+e^-\to \pi^+\pi^-$ ? \thanks{\it Dedicated to our colleague and friend Paco Yndurain who left us during the completion of the this work (to be published in the   proceedings of the QCD 08 - Montpellier conference: 7-12th july 2008).}
  } }}
  
\author{J. Layssac \thanks{Email: layssac@lpta.univ-montp2.fr} $^{\rm{a}}$ and S. Narison\thanks{Email: snarison@yahoo.fr}  \address {\footnotesize Laboratoire
de Physique Th\'eorique et Astroparticules, Universit\'e
de Montpellier II, Case 070, Place Eug\`ene
Bataillon, 34095 - Montpellier Cedex 05,
France}
}

\begin{document}

\pagestyle{myheadings}
\markright{ }
\begin{abstract}
\noindent
$C$-odd asymmetry can be studied from an accurate measurement of the angular distribution due to the interference between the $S$- and P-waves in $e^+e^-\to \pi^+\pi^-$ at order $\alpha^3$. Though the integrated total cross-section is zero as expected from the Furry's theorem, the asymmetry is dominated by the pion rescattering  diagram which is enhanced by the presence of the $\ln{s/m^2_e}$, and is quite large ($\approx$11\% at $\theta=30^0$ and $\sqrt{s} < M_{f_2}$) compared to $\alpha/\pi\simeq 0.3\%$. This process 
can also be used for alternatively measuring the size of the rescattering term and the phase of the $S$-wave amplitude, but does not help to solve the present discrepancy between the hadronic spectral functions from $e^+e^-$ and $\tau$-decay data.  
\end{abstract}
\maketitle
\section*{ Introduction}
\vspace*{-.25cm}
 \nin
\b At present, experiments on colliding electron-positron beams with high-luminosity and high-statitstics are carried out intensively.
Low-energy region below 1 GeV, in the vicinity of the $\rho,~\omega$ and $\phi$ mesons with the quantum numbers $J^{PC}=1^{- -}$, is measured with increasing precisions by the new generations of $e^+e^-$ experiments \cite{e+e-}. 
\\
\b On one hand, these precision measurements are motivated by the important contribution of this region to the anomalous magnetic moment of the muon where improved measurement is planned in the future BNL $g-2$ experiments \cite{BNL} for further tests of the Standard Model (SM) and for eventually detecting new physics beyond the SM.   At present, this project is, unfortunately, obscured by the present discrepancies between the $\pi^+\pi^-$ spectral function from $e^+e^-$ \cite{e+e-,E+E-} and $\tau$-decay data \cite{TAU}, which affects  the evaluation of the hadronic contribution to the muon anomaly  \cite{GOURDIN,CALMET,SNMU,JEGER}, and which are expected to be clarified by the future accurate data in this low-energy region.
\\
\b On  the other, they are motivated by the improved measurements of the light meson parameters which are necessary for a better understanding of the structure and of the dynamics of these mesons, which are important for testing different QCD non-perturbative methods including lattice calculations and QCD spectral sum rules predictions \cite{SNB}.
\\
\b In this paper, we shall discuss the possibility of studying properties of scalar resonances (direct coupling to $\gamma\gamma$),  and of related processes with positive $C-$parities ($\pi\pi$ non-resonant and rescattering S-wave amplitudes), in the low-energy region, which   remains a long standing puzzle in non-perturbative QCD \cite{MONTANET,SNG,MENES,PETER}, and which may eventually help for solving the $e^+e^-$ and $\tau$-decay data discrepancy.   

 \vspace*{-0.5cm}
\begin{figure}[hbt]
\begin{center}
\includegraphics[width=2.5cm]{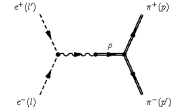}
\vspace*{-1cm}
\caption{\footnotesize Born amplitude with a $\rho$-meson-exchange in the s-channel}
\vspace*{-1cm}
\label{fig:born}
\end{center}
\end{figure}
\nin
\\
\b Contrary to the negative $C-$parities  $\rho,~\omega$ and $\phi$ mesons, which are easily observed in the processes via a one-photon exchange (Fig. \ref{fig:born}), the observation of  processes with positive $C-$parities only occurs via two-photon intermediate state (Table \ref{tab:2gamma}), which are a priori difficult to observe due to the additionnal QED coupling $\alpha^2$ suppression compared to the one in Fig. \ref{fig:born}. 
However, one may (a priori) expect that the interference between the Born amplitude in Fig. 1 and the one in Table  \ref{tab:2gamma} can be reached
at the present experimental accuracy because of the less power of $\alpha$ in the interference term relative to the diagonal one. 
\nin
\\
\b In the following, we plan to study this process,  from the  $C$-odd asymmetry angular distribution:
\beq
\vspace*{-.2cm}
{\cal A}_{ PS}\equiv\ga{d\sigma(\theta)\over d\Omega}-{d\sigma(\pi-\theta)\over d\Omega}\dr  \Big{/} {d\sigma(\theta)\over d\Omega}\Big{\vert}_{\rm Born}~,
\label{Aps}
\vspace*{-.2cm}
\eeq
which can differentiate  the $C$-even and $C$-odd parities (the Born cross-section refers to the process in Fig. \ref{fig:born}). Unlike the $C$-even process, this observable requires the detection of the charge of the final state particle, which, in the present case, is the pion. 
An analogous observable  has been discussed in the pure QED process $e^+e^-\to \mu^+\mu^-$ \cite{BELGE},
and in $e^+e^-\to f_2\to \pi^+\pi^-$ \cite{SPIN2} \footnote{We plan to reconsider this process in a future work.}.  
Here, we consider, instead, 
the interference between the usual one-photon exchange amplitude involving the $P$-wave $\pi^+\pi^-$ shown in Fig. \ref{fig:born}:
\vspace*{-.2cm}
 \beq
 e^+e^-\to \gamma \to \pi^+\pi^-~,
  \label{gamma}
  \vspace*{-.2cm}
 \eeq
with the $S$-wave amplitude shown in Table \ref{tab:2gamma}: 
\vspace*{-.2cm}
 \beq
 e^+e^-\to \gamma\gamma \to \pi^+\pi^-~,
 \label{2gamma}
 \vspace*{-.2cm}
 \eeq
 \begin{table}
\caption{\footnotesize Diagrams contributing to the process in Eq.~(\ref{2gamma})}
\label{tab:2gamma}
\vspace*{-0.25cm}
\begin{tabular}{l}
\\
\hline
\vspace*{-0.25cm}
\\
1. Contact term and Scalar-meson s-channel exchange \\   
\includegraphics[width=2.5cm]{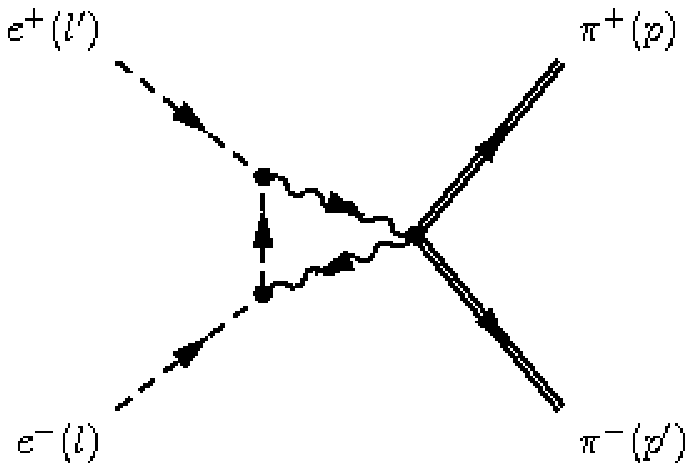}
 \includegraphics[width=2.5cm]{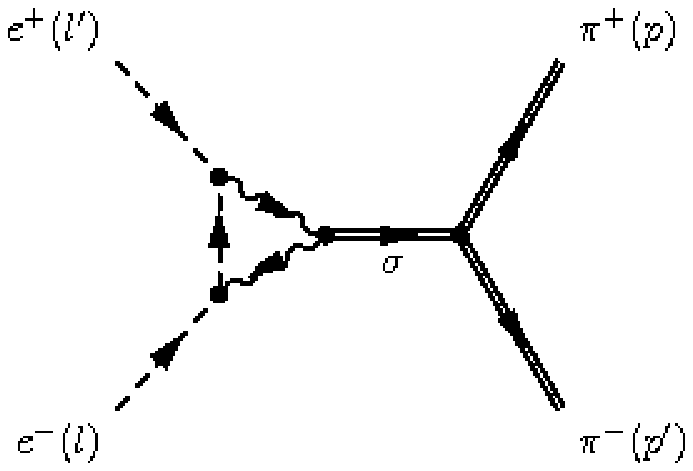}\\
\nin
 2.  One $\pi$- or $V$-meson exchange in the t channel\\
 \includegraphics[width=5cm]{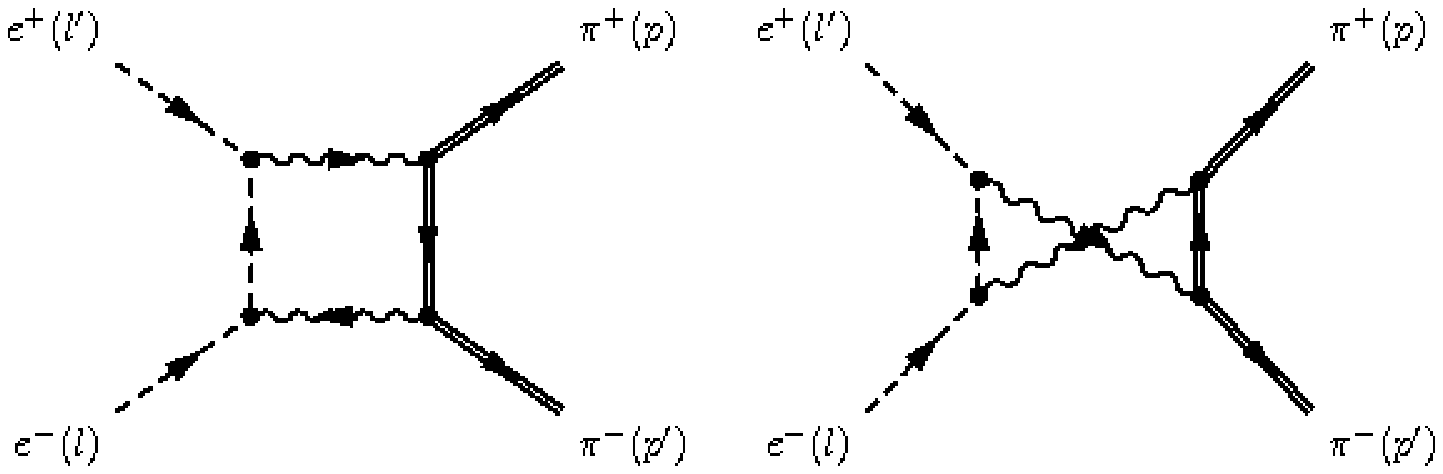}
 \vspace*{-0.3cm}
 \\
 \hline
 \vspace*{-1cm}
 \end{tabular}
\vspace*{-.3cm}
\end{table}
\nin
 which dominates over the $D$-wave one below $M_{f_2}$=1.27 GeV \footnote{This feature has been explicitly checked in $\gamma\gamma$ scattering processes \cite{MENES,PETER}.}. 
  The measurement of the $\pi^+\pi^-$ process is certainly less accurate than the pure QED final state $\mu^+ \mu^- $,
though one expects that it can be reached with  a good accuracy at the present  and forthcoming high-statistic experiments having detector with a good pion identification.  \\
\b We shall not consider the process with a real photon emission $e^+e^-\to\pi^+\pi^-\gamma$ (unless technically necessary for cancelling IR divergences), because they are usually absorbed into the radiative corrections already included into the Monte-Carlo programs or can be disentangled from the virtual photon process by tagging the real photon [the so-called initial state radiation (ISR) method used e.g. at KLOE and BABAR].
To the order at which we are working,  the only relevant of such diagrams is the interference between the amplitude where a photon is initially emitted from the lepton (ISR) with the amplitude where the photon is emitted from the pion (FSR) (the other ones are $C$-even: see e.g. \cite{e+e-}), which we shall call later on IFSR process. As emphasized in \cite{discussion}, the asymmetry of the IFSR process can be clearly measured at large photon angle and
can  be a direct measure of the ratio of the FSR over the ISR amplitude.  The KLOE measurement of this IFSR quantity agrees with Monte-Carlo simulations, where the later use a model with pointlike pions
(the pion will be also treated as a point-like particle in our calculation).
The feasibility of this measurement indicates that the alone IFSR process is an independent physical process which does not  need any additional contributions, like the ones which we shall consider in our paper, for being meaningful.  The independence of the IFSR $e^+e^-\to\pi^+\pi^-\gamma$ and of the $e^+e^-\to\pi^+\pi^-$ process considered here  will be confirmed  by the absence of IR singularities in each of them but not in their sum \footnote{This is not the case of the $e^+e^- \to \mu^+\mu^-$ and $e^+e^- \to \mu^+\mu^-\gamma$ processes calculated in Ref. \cite{e+e-}, where both contributions (real and virtual photon processes) should be added for cancelling IR divergences. We should note that the process which we shall calculate is very similar to the one : $\gamma\gamma \to \pi^+\pi^-$ which comes  from a similar box diagram. The $\gamma\gamma$ process does not also need the inclusion of real photon contribution for being meaningful.}.

   \vspace*{-.25cm}
  \section*{ Theoretical inputs }
  \vspace*{-.25cm}
  \nin
To  the order at which we are working, we use the lowest order (in 1$/f_\pi$) effective $ \pi\pi\gamma$ interaction Lagrangian:
\vspace*{-0.2cm}
\beq
{\cal L}_{\pi\pi}= ie~\Big{[}\pi^-\partial_\mu \pi^+- (\partial_\mu \pi^-)\pi^+\Big{]}A^\mu 
 + e^2 \pi^-\pi^+  A_\mu A^\mu,
\label{eq:lagrangian}
\vspace*{-.2cm}
\eeq
with $\pi$ and $A_\mu$ are the pion and photon fields. 
The scalar meson (S) couplings to $\gamma\gamma$ ($g_{S\gamma\gamma}$), to $\pi\pi (g_{S\pi\pi})$  and the pion coupling to $\gamma\gamma$ are introduced  through the interaction terms \footnote{We shall work to leading order of these couplings.}:
\vspace*{-.2cm}
\bea
{\cal L}_{S\gamma\gamma}=g_{S\gamma\gamma}SF^{(1)}_{\mu\nu}F_{(2)}^{\mu\nu}~,~~~~
 {\cal L}_{S\pi\pi}=g_{S\pi\pi}S\pi^+\pi^-~,
 \label{eq:coupling}
 \vspace*{-.2cm}
\eea
and:
\vspace*{-.2cm}
\bea
 {\cal L}_{\pi\gamma\gamma}&=&{1\over 2}g_{\pi\gamma\gamma}\pi^0\epsilon^{\mu\nu\rho\sigma}F^{(1)}_{\rho\sigma}F_{(2)}^{\mu\nu}~,
  \label{eq:coupling2}
  \vspace*{-.2cm}
 \eea
 where $F^{(i)}_{\mu\nu}$ is the photon field strength. The model dependence enters into the size of the couplings, which are normalized as in  \cite{SNG,RENARD,SNB} and fixed from the data. 
In the range of energy  where we are working, we use a vector meson dominance model (VDM) \footnote{Inclusion of resonances into chiral lagrangian has been discussed in the literature (see e.g. \cite{MENES,RAF}).}, by replacing, with a good approximation,  the virtual photon propagator by the ones of vector mesons.  This good VDM approximation being confirmed by the observation of the $\rho$-dominance of the 
$e^+e^-\to\gamma\to \pi^+\pi^-$ cross-section below 1.27 GeV.
Using VDM, the Born $s$-channel amplitude in Eq.~(\ref{gamma}) is shown in Fig.~\ref{fig:born} and reads:
\vspace*{-.2cm}
\beq
{\cal M}|_{\rm Born}=e^2~v(l'){(\hat p'-\hat p)\over (q^2+i\epsilon)} u(l) F_\pi(q^2)~,
\label{eq:ampborn}
\vspace*{-.2cm}
\eeq
where:
$
(l+l')^2=(p+p')^2=q^2\equiv s~,
$
and \footnote{We shall see later on that finite width corrections will be negligible in our calculation.}: 
\vspace*{-.2cm}
\beq
 \vert F_\pi(s)\vert^2\simeq  {M^4_\rho \ga 1+\Gamma_\rho^2/M_\rho^2\dr \over  {\vert s-M^2_\rho}-iM_\rho\Gamma_\rho\vert^2}~,
 \vspace*{-.2cm}
\eeq
is the square of pion form factor normalized as $ \vert F_\pi(s=0)\vert^2=1$.  
This expression leads to the well-known angular distribution:
\vspace*{-.2cm}
\beq
{d\sigma(\theta)\over d\Omega}\Big{\vert}_{\rm Born}\simeq \alpha^2 {|\vec{p}|^3 \over s^2\sqrt{s}}\sin^2{\theta}|F_\pi(s)|^2~,
\label{eq:angularborn}
\vspace*{-.2cm}
\eeq
where $\theta$ is the polar angle between the electron beam and the outgoing charged negative pion.
  \vspace*{-.25cm}
\section*{Evaluation of the interference amplitudes }
\vspace*{-0.25cm}
\nin
\b For this purpose, we use the chiral Lagrangian in Eq.~(\ref{eq:lagrangian}), and the couplings given previously in Eqs.~(\ref{eq:coupling}) and ~(\ref{eq:coupling2}).\\
\b We compute directly the interference term between the lowest-order Born amplitude in Eq.~(\ref{eq:ampborn}) and shown in Fig.~\ref{fig:born}, with the $\gamma\gamma$ diagrams given in Table~\ref{tab:2gamma}.  The Feynman rules for deriving the amplitude are standard. Using a Vector Meson Dominance Model (VDM), the photon line is replaced by the $\rho$-meson form factor.\\
\b Due to the complexity of the calculation, we evaluate the trace of Dirac matrices  with the $Feyncalc$ program \cite{feyncalc} linked to $Mathematica$ for expressing the results in terms of the Passarino-Veltman-'t Hooft (PAVE) integrals \cite{pave}. The analytic expressions are lengthy, which are not appropriate to present in this letter. However, they can be send by demand.\\
\b We obtain the final results by computing numerically, either with $Fortran$ or with $Mathematica$, these different PAVE integrals using the $LoopTools$ program \cite{looptools}.  \\
\b To the order we are working, we find that the contributions of the contact term and of the scalar-meson exchange in Table~\ref{tab:2gamma} are zero. These null contributions are due to the Lorentz structure of the vertex couplings and are model-independent, where the process is proportional to $l^2=l'^2=m_e^2$ as expected from symmetry arguments of the $\gamma\gamma$ amplitude.\\
\b The contributions of the last two diagrams in Table~\ref{tab:2gamma} with one pion or one vector meson exchange (within VDM) in the t-channel is UV and IR (photon mass taken to zero)  finite, which is a remarkable property. The vanishing of the UV divergence is obvious for each box diagrams due to the form of numbers of propagators and of the algebraic form of the numerators. We have checked this UV convergence by checking the absence of the $2/\epsilon$ pole or equivalently by introducing a scale $\mu_{\rm new}=e^{2/\epsilon} \mu^2_{\rm old}$.  The result is invariant when changing $\mu$ in a large range. The vanishing of the IR singularities is less trivial, which is due to a fine reorganization of the different PAVE loop integrals, and is expected in the absence of virtual photons in the t-channel~\cite{looptools}.  We have checked the absence of the IR divergence by giving a mass $\lambda$ to the photon and by varying it in a large range. As expected, the result is independent of $\lambda$. The previous successful  UV and IR numerical tests following the recommendation in the {\it LoopTools} user's manual \cite{looptools}, are a good indication on the reliability of our results, which (indirectly)indicate that, in the process which we have considered, we have not missed some other diagrams to this order.\\
\b Using the experimental value of the dominant $\omega\pi\gamma$ coupling for the t-channel vector-meson (V) exchange, one also finds that this contribution is negligible ($10^{-3}$ of  the one of pion exchange). It also indicates that the $\pi^0\pi^0$ production dominated by this contribution is  unobservable.
\vspace*{-.25cm}
\section*{Angular distribution, and C-odd Asymmetry }
\begin{figure}[hbt]
\begin{center}
\includegraphics[width=3.5cm]{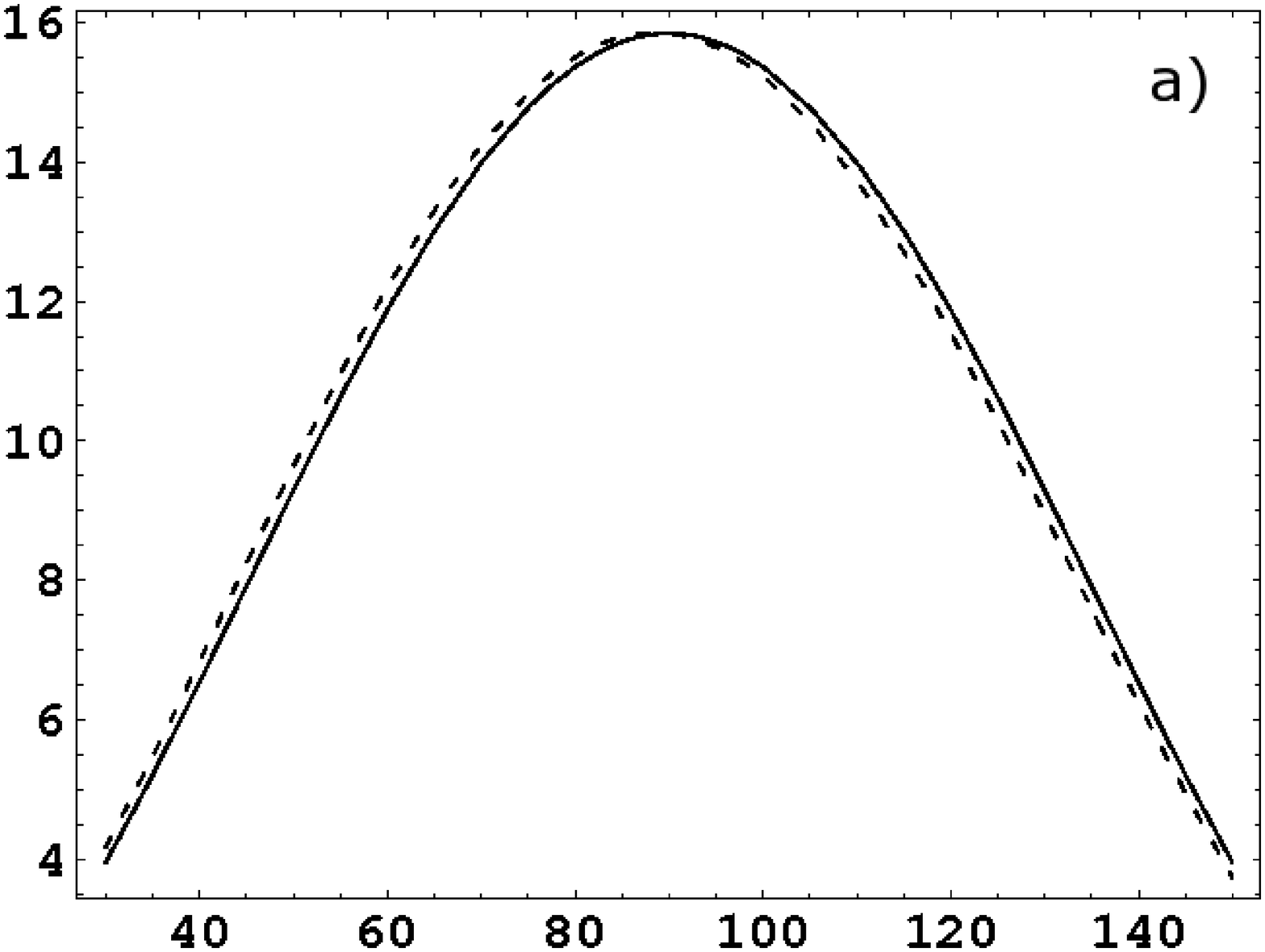}
\includegraphics[width=2.85cm]{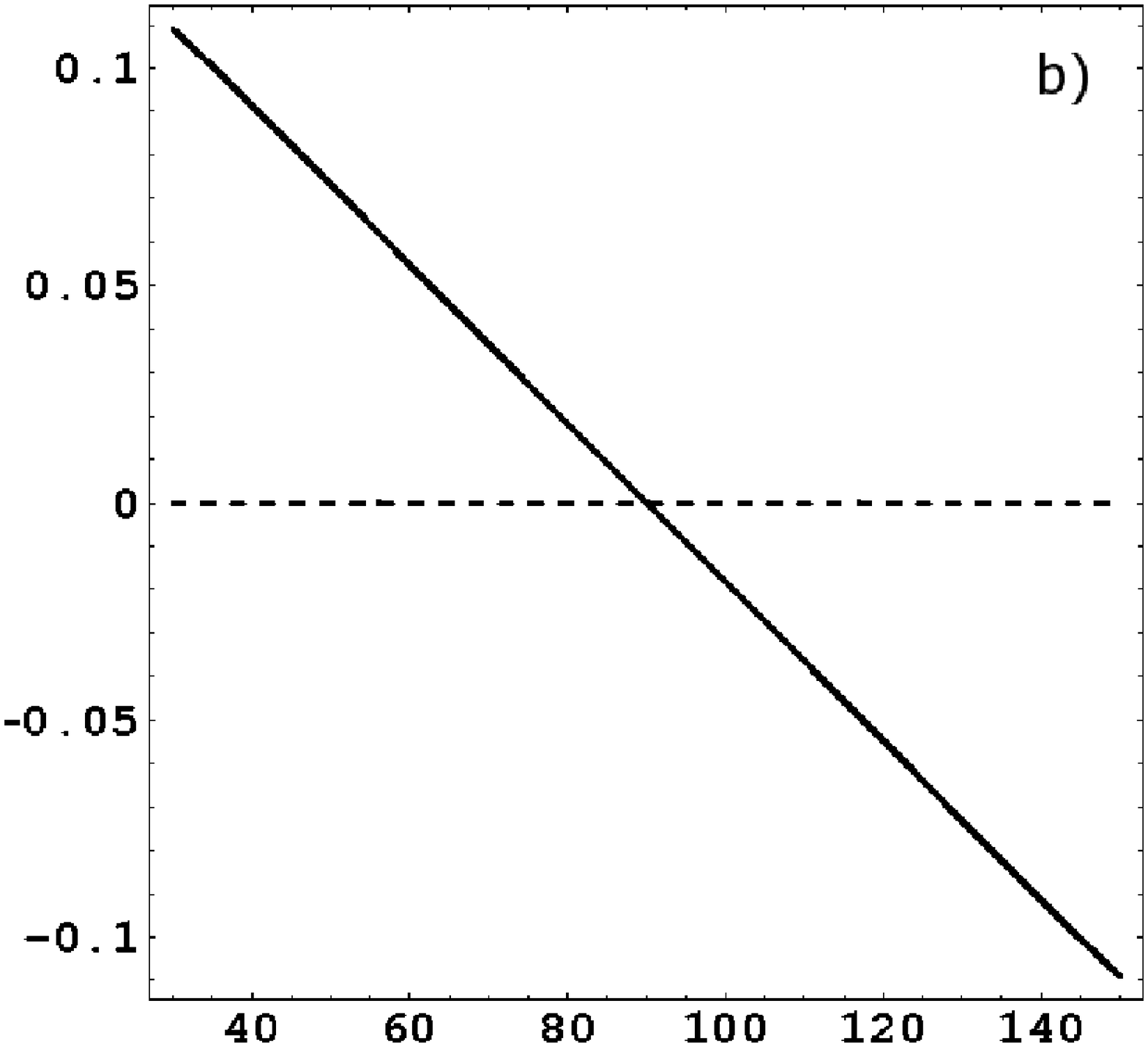}
\vspace*{-0.75cm}
\caption{\footnotesize a) Angular distribution $d\sigma / d\Omega$ in units of nanobarn versus $\theta$ (polar angle between $e^-$ and $\pi^-$) in degree
at $\sqrt{s}$=0.5 GeV: the continuous line is the Born contribution; the dashed line is
the one including radiative corrections; 
b) The asymmetry ${\cal A}_{PS}$ defined in Eq.~(\ref{Aps}) versus $\theta$ in degree. }
\label{fig:angular}
\end{center}
\vspace*{-1cm}
\end{figure}

\vspace*{-.25cm}
\nin
\b We show in Fig.~\ref{fig:angular}a) the angular distribution including radiative corrections and compared with the corresponding Born term given in Eq.~(\ref{eq:angularborn}) at the value of $\sqrt{s}=0.5$~GeV.
\vspace*{-.75cm}
\nin
The C-odd asymmetry ${\cal A}_{PS}$ defined in Eq.~(\ref{Aps}) is given in Fig.~\ref{fig:angular}b using the previous value $\sqrt{s}=0.5$~GeV.
\\
\b The radiative correction is asymmetric for $\theta$ and $\pi-\theta$, while it is maximal at small and large angles. \\
\b At $\theta=30^0$ and $\sqrt{s}$=0.5 GeV, the correction to the angular distribution is about 5.5\%, i.e. 11\% for ${\cal A}_{PS}$, which is relatively large compared with the na\"\i ve counting $(\alpha/\pi) \simeq 0.3\%$ and may be observed with improved accurate data.\\
\b We check that the $\ln^2{(s/ m^2_e)}$ contribution is zero. We fit numerically  the coefficient of the $\ln{(s/ m^2_e)}$ term expected from the electron exchange in the t-channel~\cite{looptools}. We found  the functional dependence at ${\sqrt{s}=0.5~{\rm GeV}}$ and ${\theta={\rm 30^0} }$~:
\vspace*{-.2cm}
\beq
{d\sigma\over d\Omega}\simeq {d\sigma\over d\Omega}\Big{|}_{\rm Born}\Bigg{\{}
1+\ga {\alpha\over\pi}\dr \Big{[} 2.0 \ln{s\over m^2_e}-1.5\Big{]}\Bigg{\}}~,
\label{eq:ang}
\vspace*{-.2cm}
\eeq
where the numerical coefficients contain $m_\pi^2$ and $M^2_\rho$ terms. \\
\b Eq.~(\ref{eq:ang}) shows the huge contribution from the $\ln{(s/m^2_e)}$-term. The presence of the $\ln{(s/m^2_e)}$ term seems to be a general feature of a QED calculation with a large external momentum $s$ and a virtual light particle (electron) appearing in a loop. A classical example is the QED calculation of the 2nd order correction due to electron loop for the muon $g-2$, where a $ \log (m_\mu/m_e)$ appears ($m_\mu$ is here the external momentum). 
\b Fixing again $\theta=30^0$, we study  the relative strength of the radiative corrections
 for $\sqrt{s}< M_{\rm f_2}=1.27$ GeV, where it is expected that VDM provides a good approximate
 description of the data. 
 We notice that ${\cal A}_{PS}$ is almost unaffected by the change of the $s$-values in this range of energy.
\vspace*{-.25cm}
\section*{Isospin symmetry, $\tau$-decay  and the {\it g-2} of leptons}
\vspace*{-.25cm}
\nin
The important r\^ole of the $\ln{(s/ m^2_e)}$ term present in the expression of ${\cal A}_{PS}$
[Eq. (\ref{eq:ang})] can indicate that the effect of the $C$-odd asymmetry is more pronounced in $e^+e^-\to\pi^+\pi^-$ differential cross-section than in the $\tau^- \to \nu_\tau\pi^0\pi^-$ differential decay rate.
For checking this result, we evaluate the similar process for $\tau$-decay. We found that the box diagram with internal  $\tau$, $W$ and $\gamma$ lines behaves like $\ln{(s/ M^2_\tau)}$, while the one with internal  $\nu_\tau$, $W$ and $Z$ lines vanishes in the chiral limit $m_\pi^2\to 0$. The two contributions
are relatively negligible compared to the electron case due to the $W$, $Z$ propagator suppressions
in the box diagram calculation. The difference
between the strength of the two processes indicates a violation of the isospin symmetry rotation for  the asymmetry ${\cal A}_{PS}$. 
However, the effects of the interference term vanish in the integrated  total cross-section, which is expected from the Furry theorem \footnote{As mentioned earlier, the contact term and the s-channel contributions shown in Table 1 also vanish.}. Though this result cross-checks the validity of the results obtained in this paper, it  (unfortunately) does not help to explain the present discrepancy between the hadronic spectral functions extracted from $e^+e^-\to\pi^+\pi^-$ hadron total cross-section and the one from $\tau^- \to \nu_\tau\pi^0\pi^-$ total decay rate, which are expected to be equal in the $SU(2)$ isospin symmetry limit \footnote{Some effects of I=0 scalar mesons which only contribute to $e^+e^-$ but not to $\tau$-decay have been also analyzed in \cite{SNMU}, where the contributions are tiny and do not  solve the present discrepancy between the two data.}. In fact, these spectral functions play a crucial r\^ole in the present evaluation of  the lepton anomalous magnetic $a_l\equiv 1/2(g_l-2)$ \cite{GOURDIN}--\cite{JEGER}. 
\vspace*{-.25cm}
\section*{Conclusions}
\vspace*{-.25cm}
\nin
\b We begin this paper by wondering if one can measure the $C$-odd asymmetry ${\cal A}_{PS}$ in $e^+e^-\to \pi^+\pi^-$. This project may be realized at enough small polar angle between the electron beam and outgoing  $\pi^-$ in high-statistic and hig-precision present and future experiments with a good pion identification.  \\
\b The dominance of the pion rescattering contribution indicates that contrary to the $\gamma\gamma$ and $\pi\pi$ scattering processes, it is possible  to disentangle, for this process, the pion rescattering contribution from the scalar and vector mesons exchanges. This feature being relevant in e.g. the analytic K-matrix model discussed in \cite{MENES,PETER}, where one can separate  the direct coupling of the scalar resonance to $\gamma\gamma$ from the rescattering contribution. 
The null contribution of the $\sigma$ exchange in the s-channel, i.e. of the $I=0$ part of the S-wave amplitude, may indicate that the non-zero contributions from the box diagram are only due to the $I=2$ part of the S-wave amplitude. \\
\b ${\cal A}_{PS}$ can also serve for alternatively measuring the $S$-wave phase which can be compared with the one obtained from elastic $\pi\pi$ scattering via Watson theorem.\\
\b Due to the important r\^ole of the $\ln{(s/ m^2_e)}$ term, ${\cal A}_{PS}$ due to the rescattering process may not be observed in the reaction  involving heavy leptons such as $\tau\to \nu_\tau\pi^0\pi^-$, where $m_e$ is (na\"\i vely) replaced by $M_\tau$. \\
\b Finally, the vanishing of the interference term in the total cross-section, as expected from the Furry theorem, cross-checks the validity and reliability of our results.  Unfortunately, this feature does not help to explain the present discrepancy between the hadronic spectral functions extracted from $e^+e^-\to\pi^+\pi^-$ total cross-section and the one from $\tau^- \to
\nu_\tau\pi^0\pi^-$ total decay rate which govern the hadronic contribution to the lepton anomalous magnetic moment $a_l$.
\vspace*{-.25cm}
\section*{Acknowledgements}
\vspace*{-.25cm}
\nin
We  thank A. Denig, S. Eidelman, S. Friot, G. Mennessier,  P. Minkowski, S. Mueller, E. de Rafael, F.M. Renard, G. Venenzoni and F.J. Yndurain for communications and comments.

\end{document}